\documentclass[aps,prl,twocolumn,superscriptaddress,showpacs]{revtex4-1}

\usepackage{graphicx}
\usepackage{calc}
\usepackage{bm}
\usepackage{color}

\begin{document}

\title{Specular Andreev reflection at the edge of an InAs/GaSb double quantum well with band inversion}

\author{A.~Kononov}
\affiliation{Institute of Solid State Physics RAS, 142432 Chernogolovka, Russia}
\author{S.V.~Egorov}
\affiliation{Institute of Solid State Physics RAS, 142432 Chernogolovka, Russia}
\author{V.A.~Kostarev}
\affiliation{Institute of Solid State Physics RAS, 142432 Chernogolovka, Russia}
\author{B.R.~Semyagin}
\affiliation{Institute of Semiconductor Physics, Novosibirsk 630090, Russia}
\author{V.V.~Preobrazhenskii}
\affiliation{Institute of Semiconductor Physics, Novosibirsk 630090, Russia}
\author{M.A.~Putyato}
\affiliation{Institute of Semiconductor Physics, Novosibirsk 630090, Russia}
\author{E.A.~Emelyanov}
\affiliation{Institute of Semiconductor Physics, Novosibirsk 630090, Russia}
\author{E.V.~Deviatov}
\affiliation{Institute of Solid State Physics RAS, 142432 Chernogolovka, Russia}

\date{\today}

\begin{abstract}
We experimentally investigate transport through the side junction between a niobium superconductor and the mesa edge of a two-dimensional system, realized in an InAs/GaSb double quantum well with band inversion. We demonstrate, that different transport regimes can be achieved by variation of the mesa step. We observe anomalous behavior of Andreev reflection within a finite low-bias interval, which is invariant for both transport regimes. We connect this behavior with the transition from retro- (at low biases) to specular (at high ones) Andreev reflection channels in an InAs/GaSb double quantum well with band inversion.
\end{abstract}


\maketitle

\begin{figure}
\includegraphics[width=\columnwidth]{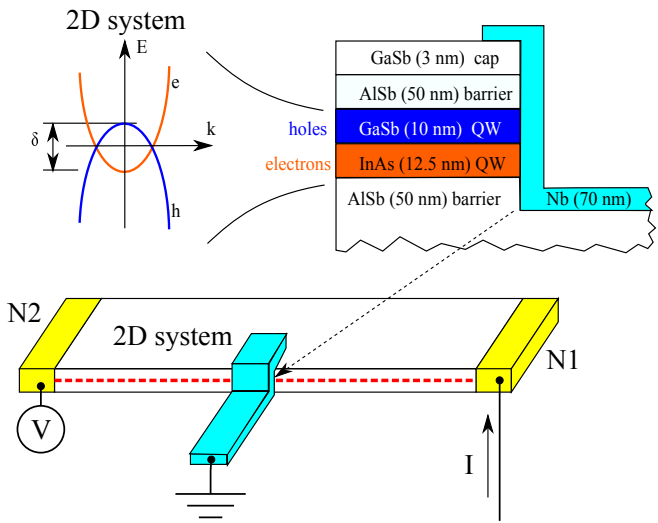}
\caption{(Color online) Sketch of the side NS  junction between a niobium superconductor and the edge of an InAs/GaSb double quantum well  with electrical connections (not in scale). The junction is fabricated  by sputtering of a thick (70~nm) Nb film (gray) over the mesa step (80~nm  or 130~nm height).
	A spectrum with an inversion gap $\delta$ is realized in an InAs/GaSb 2D system for 10~nm GaSb and 12.5 InAs wells. 
    We study charge transport across a single NS junction in a standard three-point technique: the superconducting electrode  is grounded; a current is fed through the normal Ohmic contact N1; the other normal contact (N2) traces the quantum well potential (see the main text for details).
}
\label{sample}
\end{figure}

Recent interest to an InAs/GaSb double quantum well is mostly connected with the problem of two-dimensional (2D) topological insulator~\cite{zhang1,kane,zhang2}. Similarly to the CdTe/HgCdTe quantum well~\cite{konig,kvon}, an inverted band structure can be realized in an InAs/GaSb double quantum well at some growth parameters~\cite{gasb1,gasb2,gasb3,gasb4,gasb5,gasb6}. For the 10~nm GaSb well, a spectrum with an inversion gap $\delta$ is realized for the 12~nm InAs quantum well~\cite{gasb1,gasb2,gasb3,gasb4,gasb5,gasb6}, see Fig.~\ref{sample}. If the position of the Fermi level is properly tuned by external gates,  the hybridization gap (minigap) can appear at the bands' crossings. The thinner (10~nm) or thicker (14~nm) InAs well produces a direct band 2D semiconductor or an indirect band 2D semimetal, respectively~\cite{gasb3}.   In comparison with the well-known CdTe/HgCdTe system, the InAs/GaSb double quantum well provides the stability of a III-V material and well-developed preparation  technology.

Different correlated systems with band inversion are expected to demonstrate non-trivial physics in  proximity with a superconductor. For the topological insulators~\cite{zhang1,kane,zhang2},  it  allows topological superconductivity  regime~\cite{Fu,yakoby}, which stimulates a search for  Majorana fermions~\cite{reviews}.  In the case of a Weyl semimetal~\cite{weyl}, the proximity is predicted~\cite{spec} to produce  specular Andreev reflection~\cite{been1,been2}.

Andreev reflection~\cite{andreev} allows charge transport from normal metal (N) to superconductor (S) at energies below the superconducting gap. An electron is injected through the NS interface by creating  a Cooper pair, so a hole is reflected back to the N side of the junction~\cite{andreev,tinkham}. Usually, the reflected hole remains in the conduction band of the normal metal (so called retro-, or intraband, Andreev reflection - RAR)~\cite{andreev}. However, for some specific situations,  a hole can appear in the valence band, which is known as specular (or interband) Andreev reflection (SAR)~\cite{been1,been2}. The latter has been recently reported for graphene~\cite{efetov}.

For some superconducting or ferromagnetic metals,  a junction with 2D systems can  be conveniently realized as a side junction at the mesa step~\cite{nbsemi,nbhgte,feinas}. A side superconducting contact is primary connected to the 2D edge, which transport properties are defined by the edge potential~\cite{shklovskii,image02}. Since the Andreev reflection is strongly affected by the scattering at the NS interface~\cite{BTK},  different transport regimes can be  achieved by variation of the edge potential strength, e.g. by variation of the mesa step~\cite{kozlov}.

Here, we experimentally investigate transport through the side junction between a niobium superconductor and the mesa edge of a two-dimensional system, realized in an InAs/GaSb double quantum well with band inversion. We demonstrate, that different transport regimes can be achieved by variation of the mesa step. We observe anomalous behavior of Andreev reflection within a finite low-bias interval, which is invariant for both transport regimes. We connect this behavior with the transition from retro- (at low biases) to specular (at high ones) Andreev reflection channels in an InAs/GaSb double quantum well with band inversion.

Our samples are grown by solid source molecular beam epitaxy on semi-insulating GaAs (100)  substrates. The active layer is composed of two,  12.5-nm thick InAs and 10-nm thick GaSb,  quantum wells, sandwiched between two 50-nm thick AlSb barriers. Details on the growth parameters can be found elsewhere~\cite{growth}. As obtained from standard magnetoresistance measurements, the 2D system is characterized by bulk electron-type conductivity. The mobility at 4K is about $2 \cdot 10^{4}  $cm$^{2}$/Vs  and the carrier density is   $2 \cdot 10^{12}  $cm$^{-2}$.

A sample sketch is presented in Fig.~\ref{sample}. The 100~$\mu$m wide mesa is formed by wet chemical etching. To  realize different transport regimes at the NS interface, the samples differ by the mesa step height. Shallow etching (80 nm) is stopped just after the bottom InAs quantum well.  The neighbor AlSb barrier is also removed for samples with deep etching (130~nm). We suppose, that variation of the mesa step height leads to variation of the edge potential strength~\cite{kozlov}. 

We fabricate side contacts to  an InAs/GaSb double quantum well by coating  the mesa step by a metallic film~\cite{nbsemi,feinas} with some overlap (2-3~$\mu$m). Because of the insulating upper AlSb layer, see Fig.~\ref{sample}, the side contact to a 2D system is independent of the exact overlap value. We fabricate two Ohmic contacts by thermal evaporation of 100~nm Au (with few nm Ni to improve adhesion). These Ohmic contacts are characterized by a constant, bias-independent, ($\approx 1 k\Omega$) resistance. To prepare  superconducting NS junctions, we use dc sputtering to deposit a 70~nm thick  Nb  film at the  mesa step. The 10~$\mu$m wide Nb electrodes are formed by lift-off technique.  To avoid mobility  degradation, the sample is kept at room temperature during the sputtering process.

	We study charge transport across a single NS (i.e. 2D -- Nb) junction in a standard three-point technique, see Fig.~\ref{sample}: the superconducting electrode  is grounded; a current is fed through one of the normal Ohmic contacts, N1 in Fig.~\ref{sample}; the other normal contact (N2, respectively) traces the quantum well potential. 

In a three-point technique, the measured potential $V$ reflects in-series connected resistances of the grounded contact and the 2D system. In our experiment the former term is dominant, because of the relatively low in-plane 2D resistance (about 100 $\Omega$ at present concentration and mobility). In this case, the 2D edge is   equipotential, so the measured $V$ reflects the behavior of the particular (grounded) NS   interface, since the superconducting Nb electrode is of zero resistance.  To support this conclusion experimentally, the obtained $I-V$ characteristics are verified to be independent of the mutual positions of  current and voltage probes. 

We sweep a dc current component from -2 to +2~$\mu$A. To obtain $dV/dI(V)$ characteristics,   this dc current is additionally modulated by a low ac (30~pA, 110~Hz) component. We measure both,  dc ($V$) and ac ($\sim dV/dI$), components of the double quantum well potential by using a dc voltmeter and a lock-in, respectively. We check, that the lock-in signal is independent of the modulation frequency in the 60~Hz -- 300~Hz range, which is defined by applied ac filters.  To extract features specific to an InAs/GaSb system, the measurements are performed at a temperature of 30~mK. Similar results are obtained from different samples in several cooling cycles.


\begin{figure}
\includegraphics[width=\columnwidth]{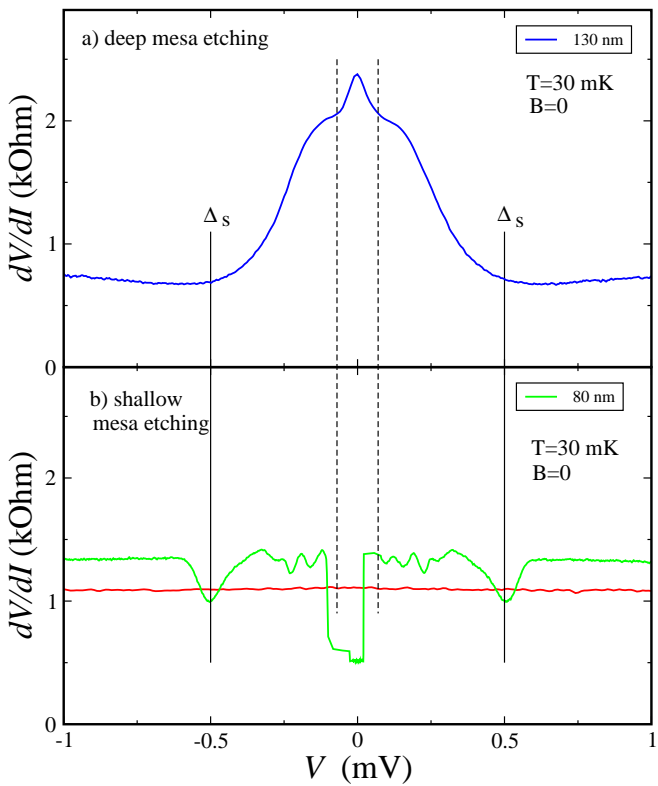}
\caption{(Color online) 
Fig.~\ref{IV} Examples of  $dV/dI(V)$ characteristics of a single NS junction  for deep (a) and shallow (b) mesa etching. In both cases,  the superconducting gap $\Delta_{s}\simeq\pm$~0.5~mV (denoted by thin solid lines) can be clearly identified. The subgap resistance  is undoubtedly finite, which is only possible due to Andreev reflection.  In the case of deep etching, (a), single-particle scattering is significant at the NS interface, since the subgap resistance strongly exceeds the normal (out-of-the-gap) value~\cite{BTK}. In the case of shallow etching, (b),  the NS interface is much more transparent. The linear Ohmic behavior of a normal Au contact is demonstrated by red $dV/dI(V)$ curve.   Specifics of the InAs/GaSb double quantum well seems to  appear in the narrow $\pm$~0.07~mV bias interval around  zero, denoted by two thin dashed lines. The full width of this interval 0.14~mV$<<\Delta_s$ is identical in (a) and (b) cases. The bias interval is asymmetric in the (b) case because of hysteresis with sweep direction at low biases.   All the curves are obtained at  the temperature $T=30$~mK$<<T_c$ in zero magnetic field. 
} 
\label{IV}
\end{figure}

Fig.~\ref{IV} presents the examples of  $dV/dI(V)$ characteristics of a single SN junction  for deep (a) and shallow (b) mesa etching. 
In both cases,  the superconducting gap $\Delta_{s}\simeq\pm$~0.5~mV (denoted by thin solid lines) can be clearly identified. For our Nb films, $\Delta_{s}$ is reduced in comparison with the bulk Nb value $\Delta_{Nb}\simeq\pm$~1.15~mV, because of non-perfect sputtering environment. The subgap resistance  is undoubtedly finite, which is only possible due to Andreev reflection~\cite{andreev,tinkham}.

Qualitative difference between $dV/dI(V)$ curves in Fig.~\ref{IV} (a) and (b) supports our initial idea of different transport regimes at the NS interface. The edge potential is stronger for deeper mesa etching~\cite{kozlov}, which results in different scattering regimes~\cite{BTK} of Andreev reflection:

(i) In the case of deep etching, see Fig.~\ref{IV} (a), the subgap resistance $R_{max}\approx 2.0$~k$\Omega$  exceeds the normal junction resistance $R_{N}\approx 0.75$~k$\Omega$, so  single-particle scattering is significant at the Nb-2D interface. A transmission of the  interface $T$ can be estimated as $\approx 0.37$, which corresponds to the BTK barrier strength~\cite{BTK} $Z\approx 1.3$. 

(ii) In the case of shallow etching, the Nb-2D interface is much more transparent, see Fig.~\ref{IV} (b), so the interface transmission $T$ is closer to 1.  

Specifics of the InAs/GaSb 2D system seems to  appear in the narrow $\pm$~0.07~mV bias interval around  zero, which is denoted by two dashed lines in Fig.~\ref{IV}. In the case of deep mesa etching, (a), the subgap resistance is increased within this bias interval. In the case of shallow mesa etching, (b), the subgap resistance drops strongly below the normal resistance level $R_{N}\approx 1.3$~k$\Omega$ within the bias interval of the same, 0.14~mV$<<\Delta_s$, width. Andreev reflection is enhanced to the ideal, scattering-free, regime~\cite{andreev,tinkham,BTK}. The bias interval is slightly asymmetric  in the (b) case because of hysteresis with sweep direction at low biases.  
Fig.~\ref{IV} (b) also demonstrates the linear Ohmic behavior of a normal Au contact. There is no any specific, bias-dependent behavior in this case. 

\begin{figure}
\includegraphics[width=\columnwidth]{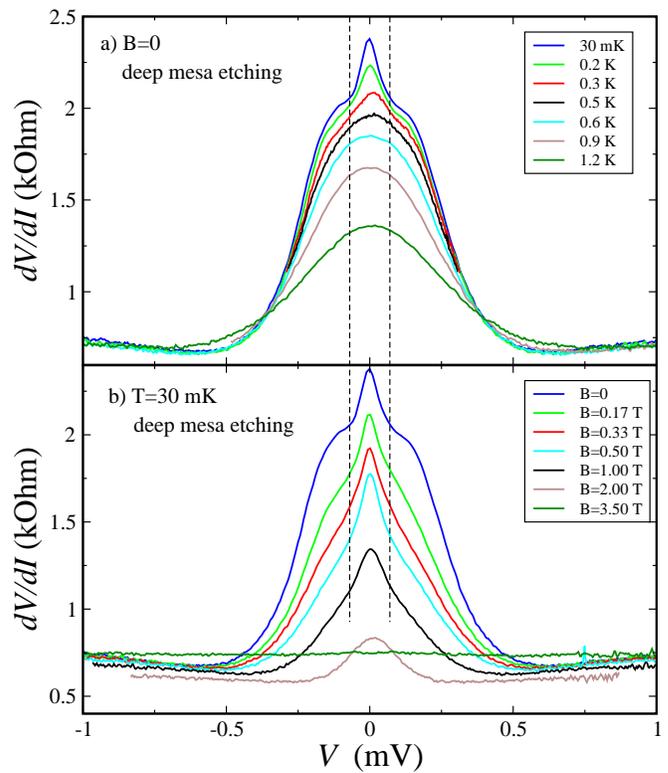}
\caption{(Color online) Superconductivity suppression  by temperature (a) or in-plane magnetic field (b) for the deep mesa sample. The temperature 1.2~K$<<T_c$ is not enough for  complete  suppression (a), but the $dV/dI(V)$ curve is linear above the critical magnetic field $B_c=2.1$~T (b). The specific resistance peak at low biases disappears earlier: it can not be seen above 0.5~K temperature and 1~T magnetic field.
} 
\label{IV_BT}
\end{figure}

Fig.~\ref{IV_BT} demonstrates the superconductivity suppression  by temperature (a) or in-plain magnetic field (b) for the deep mesa sample. The temperature range 30~mK-1.2~K is not enough for complete temperature suppression (a), but the $dV/dI(V)$ curve is linear above the critical magnetic field 2.1 T (b). The specific resistance peak at low biases disappears earlier: it can not be seen above 0.5~K temperature and 1~T magnetic field, see  Fig.~\ref{IV_BT} (a) and (b), respectively. 

\begin{figure}
\includegraphics[width=\columnwidth]{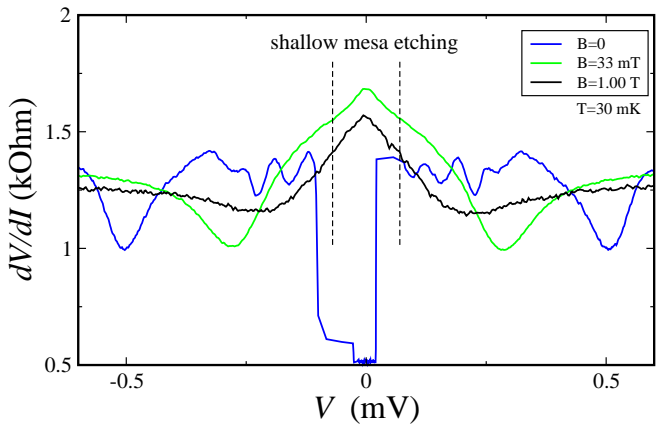}
\caption{(Color online) Switching of the $dV/dI(V)$ qualitative behavior by low ($\approx 33$~mT) magnetic field for the shallow mesa sample: the resistance drop is converted into the resistance peak within the same $\pm$~0.07~mV bias interval. Similarly to the deep mesa sample, the peak disappears completely above 1~T.  All the curves are obtained at the minimal $T=30$~mK temperature.} 
\label{B_switching}
\end{figure}

Similar temperature suppression is observed for the shallow mesa sample, but the magnetic field dependence is more complicated. Fig.~\ref{B_switching} demonstrates that low ($\approx 33$~mT) magnetic field switches the $dV/dI(V)$ qualitative behavior to one, characteristic for the deep etching case: the resistance drop is converted into the resistance peak within the same $\pm$~0.07~mV bias interval. Similarly to the deep mesa sample, the peak disappears completely above 1~T.

Our experiment essentially demands some non-trivial physical explanation:

(i) In general, any specifics of the normal-side spectrum,  like Zeeman~\cite{tinkham} or spin-orbit splitting~\cite{inoue},  can not lead to any low-energy effects in usual (retro-) Andreev reflection. The only relevant energy scale is the superconducting gap~\cite{tinkham}.

(ii) The behavior of the subgap resistance can not be connected with the disorder. The latter can only provide a small, weak antilocalization-like correction at zero bias, known as disorder-enhanced Andreev reflection~\cite{wal1,wal2}. In contrast, the low-bias resistance is seriously increased in Figs.~\ref{IV} (a) and drops twice below the normal junction's value in Fig.~\ref{IV} (b). Also, the bias interval is finite and independent of the scattering at the interface. 
 
We believe that this anomalous subgap resistance behavior corresponds to a transition from retro- to specular Andreev reflection channels, which is defined by the inversion gap $\delta$ in the 2D spectrum of an InAs/GaSb double-well system.

The bulk of the sample is characterized by electron-type conductivity, i.e. the hole subband is empty in Fig.~\ref{sample}.   By approaching the edge, the electron concentration is  diminishing to zero, because the sample edge is a potential barrier for  carriers.  The transport (scattering) regime of the NS interface is defined by the edge potential strength: in the case  of shallow etching the carriers' concentrations are finite at the interface, while the deep mesa step is characterized by a depletion region of finite width~\cite{shklovskii}.  The edge electrostatics demands a smooth edge potential profile, i.e. gradual diminishing of the Fermi level position~\cite{shklovskii}. Thus, the Fermi level is necessarily placed within the inversion spectral gap $\delta$ near the sample edge. For this reason,  both electron and hole spectrum branches in Fig.~\ref{sample} can contribute to transport at the NS interface.

If the Fermi level is placed within the inversion gap $\delta$ at the sample edge, see Fig.~\ref{sample}, usual intraband (retro-) Andreev reflection  is only possible for biases $eV$ below the inversion gap $\delta/2$. At high biases $\delta/2<eV<\Delta_{s}$, the reflected hole should necessarily belong to the other spectrum branch, which results in specular  reflection at the interface. (In principle, SAR is also allowed at low biases, but RAR is completely prohibited at $\delta/2<eV<\Delta_{s}$.) Thus, a transition from retro- to specular Andreev reflection channels should appear at $eV=\delta/2$ bias. 

 This is exactly that we obtain in the experiment, see Fig.~\ref{IV}: at  biases around 0.07~mV, drastic change in the subgap resistance occurs for both scattering regimes at the NS interface. This bias can therefore be identified as a crossover point between the intra-(RAR) and inter-(SAR) band channels. In the case of the transparent NS interface, switching off the RAR channel is accompanied by increase in the differential resistance. This crossover is subjected to some hysteresis, see Fig.~\ref{IV} (b).  

Depletion at the NS interface in the deep etching case, see in Fig.~\ref{IV} (a), partially suppresses both RAR~\cite{BTK} and SAR~\cite{been1,been2} channels. From Fig.~\ref{IV} (a) we can conclude, that SAR is less sensitive to the single-particle scattering at the interface, e.g. because of different transmission probabilities for two spectrum branches in  Fig.~\ref{sample}.

This picture is consistent with the temperature and  magnetic field dependencies in Fig.~\ref{IV_BT} and  Fig.~\ref{B_switching}:  

(i) Even low magnetic field induces depletion at the 2D edge~\cite{shklovskii}. The transparent transport regime is converted into the scattering-dominant (like in the deep mesa case). The resistance drop is therefore converted into the resistance peak within the same $\pm$~0.07~mV bias interval as for the deep etching case in Fig.~\ref{B_switching}.

(ii) The magnetic field suppression of Andreev reflection is different for the intraband (RAR, low-bias) and the interband (SAR, high-bias)  channels for the band overlap spectrum in Fig.~\ref{sample}. RAR should be fully suppressed if the Zeeman splitting $\Delta_Z$ exceeds the inversion gap  $\delta$, while the specular process survives up to much higher $\Delta_Z \sim \Delta_s$. Similar considerations can be applied to the temperature smearing. If we estimate $\delta/2=0.07$~meV from the experiment, the low-bias anomaly should disappear at $kT=\delta/2 \approx 0.5$~K$<<T_c$ or in the magnetic field $B\approx B_c \times  \delta / \Delta_s \approx 0.6$~T. 

These estimations are consistent with the experiment, see Fig.~\ref{IV_BT}, the estimation $\delta=0.14$~meV is also consistent with the value reported in Ref.~\cite{gasb3}. Moreover, similar low-bias behavior has been reported~\cite{nbsemi} for another 2D system with band overlap,  the 20~nm CdTe/HgCdTe quantum well, which supports the proposed physical picture. It is worth mentioning, that the crossover from SAR to RAR  has also been identified in Ref.~\cite{efetov} for graphene by the drastic change in the subgap resistance.

We wish to thank  A.M.~Bobkov, I.V.~Bobkova, Ya.~Fominov, V.T.~Dolgopolov, and T.M.~Klapwijk for fruitful discussions.  We gratefully acknowledge financial support by the RFBR (project No.~16-02-00405), RAS and the Ministry of Education and Science of the Russian Federation under Contract No. 14.B25.31.0007.

\end{document}